# Data Diffusion: Dynamic Resource Provision and Data-Aware Scheduling for Data-Intensive Applications

Ioan Raicu[1], Yong Zhao[2], Ian Foster[1,3,4], Alex Szalay[5]

*iraicu@cs.uchicago.edu, yozha@microsoft.com, foster@mcs.anl.gov, szalay@jhu.edu*

[1]Department of Computer Science, University of Chicago, IL, USA

[2]Microsoft Corporation, Redmond, WA, USA

[3]Computation Institute, University of Chicago and Argonne National Laboratory, USA

[4]Mathematics and Computer Science Division, Argonne National Laboratory, Argonne IL, USA

[5]Department of Physics and Astronomy, The Johns Hopkins University, Baltimore MD, USA

## ABSTRACT

*Data intensive applications often involve the analysis of large datasets that require large amounts of compute and storage resources. While dedicated compute and/or storage farms offer good task/data throughput, they suffer low resource utilization problem under varying workloads conditions. If we instead move such data to distributed computing resources, then we incur expensive data transfer cost. In this paper, we propose a data diffusion approach that combines dynamic resource provisioning, on-demand data replication and caching, and data locality-aware scheduling to achieve improved resource efficiency under varying workloads. We define an abstract "data diffusion model" that takes into consideration the workload characteristics, data accessing cost, application throughput and resource utilization; we validate the model using a real-world large-scale astronomy application. Our results show that data diffusion can increase the performance index by as much as 34X, and improve application response time by over 506X, while achieving near-optimal throughputs and execution times.*

## 1. INTRODUCTION

The ability to analyze large quantities of data has become increasingly important in many fields. To achieve rapid turnaround, data may be distributed over hundreds of computers. In such circumstances, data locality has been shown to be crucial to the successful and efficient use of large distributed systems for data-intensive applications [7, 29].

One approach to achieving data locality—adopted, for example, by Google [3, 10]—is to build large compute-storage farms dedicated to storing data and responding to user requests for processing. However, such approaches can be expensive (in terms of idle resources) if load varies significantly over the two dimensions of time and/or the data of interest.

We previously outline [31] an alternative *data diffusion* approach, in which resources required for data analysis are acquired dynamically, in response to demand. Resources may be acquired either "locally" or "remotely"; their location only matters in terms of associated cost tradeoffs. Both data and applications are copied (they "diffuse") to newly acquired resources for processing. Acquired resources (computers and storage) and the data that they hold can be "cached" for some time, thus allowing more rapid responses to subsequent requests. If demand drops, resources can be released, allowing their use for other purposes. Thus, data diffuses over an increasing number of CPUs as demand increases, and then contracting as load reduces. We have implemented the data diffusion concept in Falkon, a Fast and Light-weight tasK executiON framework [4, 11].

Data diffusion involves a combination of dynamic resource provisioning, data caching, and data-aware scheduling. The approach is reminiscent of cooperative caching [16], cooperative web-caching [17], and peer-to-peer storage systems [15]. (Other data-aware scheduling approaches tend to assume static resources [1, 2].) However, in our approach we need to acquire dynamically not only storage resources but also computing resources. In addition, datasets may be terabytes in size and data access is for analysis (not retrieval). Further complicating the situation is our limited knowledge of workloads, which may involve many different applications.

In principle, data diffusion can provide the benefit of dedicated hardware without the associated high costs. It can also overcome inefficiencies that arise when executing data-intensive applications in distributed environments, due to high costs of data movement [29]: if workloads have sufficient internal locality of reference [20], then it is feasible to acquire and use remote resources despite high initial data movement costs.

The performance achieved with data diffusion depends crucially on the characteristics of application workloads and the underlying infrastructure. As a first step towards quantifying these dependences, one of our previous studies [31] conducted experiments with both micro-benchmarks and a large scale astronomy application, and showed that data diffusion improves performance relative to alternative approaches, and provides improved scalability and aggregated I/O bandwidth.

Our previous results did not consider the dynamic resource provisioning aspect of data diffusion. This paper's focus is to explore the effects of provisioning on application performance, a central theme in data diffusion. We also introduce here an abstract model that formally defines data diffusion, and which can be used to study its effects in different scenarios at a theoretical level. Finally, we perform a preliminary model validation study on results from a real large-scale astronomy application. [6, 31]

## 2. RELATED WORK

The results presented here build on our past work on resource provisioning [11] and task dispatching [4], and data diffusion [23, 31]. This section is partitioned in two, first covering related work in resource provisioning (i.e. multi-level scheduling) and then data management.

Multi-level scheduling has been applied at the OS level [27, 30] to provide faster scheduling for groups of tasks for a specific user or purpose by employing an overlay that does lightweight scheduling within a heavier-weight container of resources: e.g., threads within a process or pre-allocated thread group.

Frey et al. pioneered the application of this principle to clusters via their work on Condor "glide-ins" [35]. Requests to a batch scheduler (submitted, for example, via Globus GRAM4 [34]) create Condor "startd" processes, which then register with a Condor resource manager that runs independently of the batch scheduler. Others have also used this technique. For example, Mehta et al. [38] embed a Condor pool in a batch-scheduled cluster, while MyCluster [36] creates "personal clusters" running Condor or SGE. Such "virtual clusters" can be dedicated to a single workload. Thus, Singh et al. find, in a simulation study [37], a reduction of about 50% in completion time, due to reduction in queue wait time. However, because they rely on heavyweight schedulers to dispatch work to the virtual cluster, the per-task dispatch time remains high.

In a different space, Bresnahan et al. [41] describe a multi-level scheduling architecture specialized for the dynamic allocation of compute cluster bandwidth. A modified Globus GridFTP server varies the number of GridFTP data movers as server load changes.

Appleby et al. [39] were one of several groups to explore dynamic resource provisioning within a data center. Ramakrishnan et al. [40] also address adaptive resource provisioning with a focus primarily on resource sharing and container level resource management. Our work differs in its focus on resource provisioning on non-dedicated resources managed by local resource managers (LRMs).

Shifting our focus to data management, we believe coupling it with resource management will be most effective. Ranganathan et al. used simulation studies [9] to show that proactive data replication can improve application performance. The Stork [25] scheduler seeks to improve performance and reliability when batch scheduling by explicitly scheduling data placement operations. However, while Stork can be used with other system components to co-schedule CPU and storage resources, there is no attempt to retain nodes between tasks as in our work.

The GFarm team implemented a data-aware scheduler in Gfarm using an LSF scheduler plugin [1, 21]. Their performance results are for a small system (6 nodes, 300 jobs, 900 MB input files, 2640 second workload without data-aware scheduling, 1650 seconds with data-aware scheduling, 0.1–0.2 jobs/sec, 90MB/s to 180MB/s data rates); it is not clear that it scales to larger systems. In contrast, we have tested our proposed data diffusion with 75 nodes, 250K jobs, input data ranging from 1B to 1GB, workflows exceeding 1000 jobs/sec, and data rates exceeding 8750 MB/s. [31]

BigTable [19], Google File System (GFS) [3], and MapReduce [10] (as well as Hadoop [24]) couple data and computing resources to accelerate data-intensive applications. However, these systems all assume a static set of resources. Furthermore, the tight coupling of execution engine (MapReduce, Hadoop) and file system (GFS) means that applications that want to use these tools must be modified. In our work, we further extend this fusion of data and compute resource management by also enabling dynamic resource provisioning, which we assert can provide performance advantages when workload characteristics change over time. In addition, because we perform data movement prior to task execution, we are able to run applications unmodified.

The batch-aware distributed file system (BAD-FS) [26] caches data transferred from centralized data storage servers to local disks. However, it uses a capacity-aware scheduler which is differentiated from a data-aware scheduler by its focus on ensuring that jobs have enough capacity to execute, rather than on placing jobs to minimize cache-to-cache transfers. We expect BAD-FS to produce more local area traffic than data diffusion. Although BAD-FS addresses dynamic deployment via multi-level scheduling, it does not address dynamic reconfiguration during the lifetime of the deployment, a key feature offered in Falkon, and essential in achieving good resource efficiency in time-varying load workloads.

# 3. DATA DIFFUSION ARCHITECTURE

We describe the practical realization of data diffusion in the context of the Falkon task dispatch framework [4, 31]. We also discuss the data-aware scheduler design, algorithm, and various policies.

## 3.1 Falkon and Data Diffusion

To enable the rapid execution of many tasks on distributed resources, Falkon combines (1) multi-level scheduling [12, 13] to separate resource acquisition (via requests to batch schedulers) from task dispatch, and (2) a streamlined dispatcher to achieve several orders of magnitude higher throughput (487 tasks/sec) and scalability (54K executors, 2M queued tasks) than other resource managers [4]. Recent work has achieved throughputs in excess of 3750 tasks/sec and the management of up to 1M simulated executors without significant degradation of throughput. [32]

The Falkon architecture comprises a set of (dynamically allocated) *executors* that cache and analyze data; a *dynamic resource provisioner* (DRP) that manages the creation and deletion of executors; and a *dispatcher* that dispatches each incoming task to an executor. The provisioner uses tunable allocation and de-allocation policies to provision resources adaptively. Individual executors manage their own caches, using local eviction policies, and communicate changes in cache content to the dispatcher. The dispatcher sends tasks to nodes that have cached the most needed data, along with the information on how to locate needed data. An executor that receives a task to execute will, if possible, access required data from its local cache or request it from peer executors. Only if no cached copy is available does the executor request a copy from persistent storage.

### *3.1.1 Data Diffusion Architecture*

To support location-aware scheduling, we implement a centralized index within the dispatcher that records the location of every cached data object. This index is maintained loosely coherent with the contents of the executor's caches via periodic update messages generated by the executors. In addition, each executor maintains a local index to record the location of its cached data objects. We believe that this hybrid architecture provides a good balance between latency to the data and good scalability; see our previous work [31] for a deeper analysis in the difference between a centralized index and a distributed one, and under what conditions a distributed index is preferred.

Figure 1 shows the Falkon architecture, including both the data management and data-aware scheduler components. We start with a user which submits tasks to the Falkon wait queue. The wait queue length triggers the dynamic resource provisioning to allocate resources via GRAM4 [34] from the available set of resources, which in turn allocates the resources and bootstraps the executors on the remote machines. The black dotted lines represent the scheduler sending the task to the compute nodes, along with the necessary information about where to find input data. The red thick solid lines represent the ability for each executor to get data from remote persistent storage. The blue thin solid lines represent the ability for each storage resource to obtain cached data from another peer executor. The current implementation runs a GridFTP server [30] alongside each executor, which allows other executors to read data from its cache.

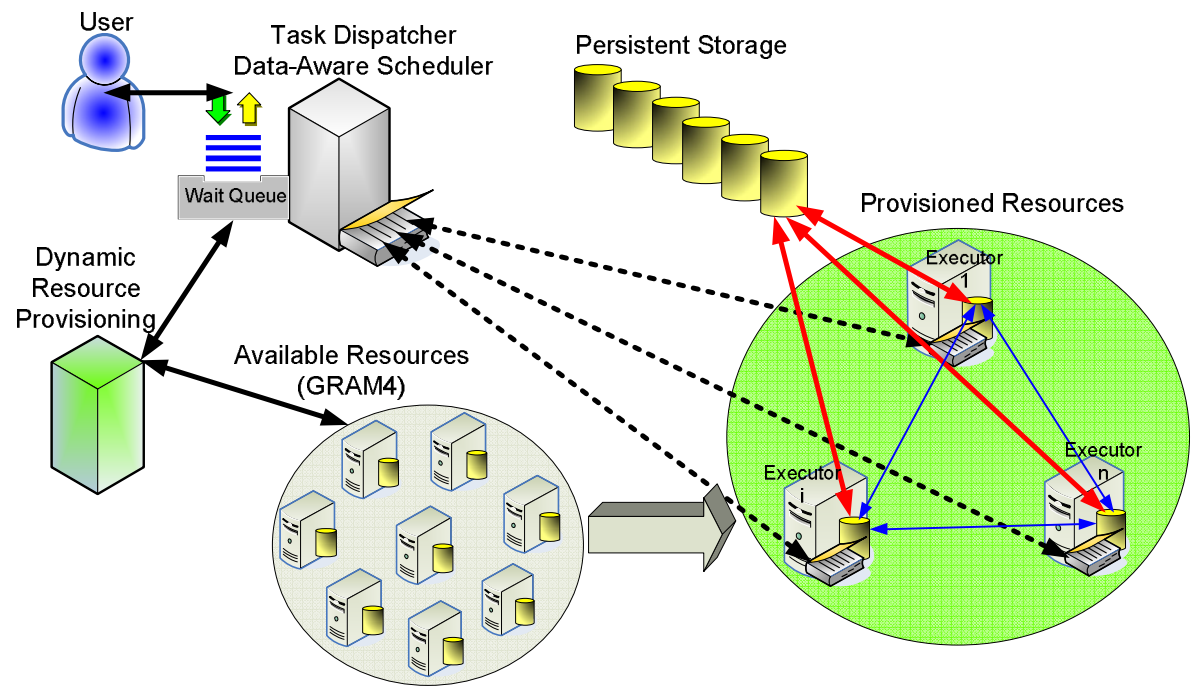

***Figure 1: Architecture overview of Falkon extended with data diffusion (data management and data-aware scheduler)***

We assume that data is not modified after initial creation, an assumption that we found to be true for many data analysis applications. Thus, we can avoid complicated and expensive cache coherence schemes. We implement four well-known cache eviction policies [16]: *Random*, *FIFO* (First In First Out), *LRU* (Least Recently Used), and *LFU* (Least Frequently Used). The experiments in this paper all use LRU; we will study the effects of other policies in future work.

## 3.2 Data-Aware Scheduler Design

The data-aware scheduler is central to the success of data diffusion, as harnessing the data-locality from application access patterns is crucial to achieving good performance and scalability for data-intensive applications. This section covers the data-aware scheduler and the parameters that affect its performance.

We implement five task dispatch policies: 1) first-available, 2) first-cache-available, 3) max-cache-hit, 4) max-compute-util, and 5) good-cache-compute [27, 31]. We omit to discuss policy (2) as it does not have any advantages over the other policies in practice.

The **first-available** policy ignores data location information when selecting an executor for a task; it simply chooses the first available executor, and furthermore provides the executor with no information concerning the location of data objects needed by the task. Thus, the executor must fetch all data needed by a task from persistent storage on every access. This policy is used for all experiments that do not use data diffusion.

The **max-cache-hit** policy uses information about data location to dispatch each task to the executor with the largest number of data needed by that task. If that executor is busy, task dispatch is delayed until the executor becomes available. This strategy can be expected to reduce data movement operations compared to first-cache-available and max-compute-util, but may lead to load imbalances where CPU utilization will be sub optimal, especially if data popularity is not uniform or nodes frequently join and leave (i.e. this is the case for dynamic resource provisioning under varying loads). This policy is most suitable for data-intensive workloads.

The **max-compute-util** policy also leverages data location information. This policy attempts to maximize the resource utilization even at the potential higher cost of data movement. It always sends a task to an available executor, but if there are several candidates, it chooses the one that has the most data needed by the task. This policy is most suitable for compute-intensive workloads.

We believe that a combination of policy (3) and (4) will lead to good results in practice, as we also show in the performance evaluation in this paper. We have two heuristics to combine these two policies, into a new policy called **good-cache-compute**, which attempts to strike a good balance between these two policies. The first heuristic is based on the CPU utilization, which sets a threshold to decide when to use policy (3) and when to use policy (4). A value of 90% works well in practice as it keeps CPU utilization above 90% and it gives the scheduler some flexibility to improve the cache hit rates significantly when compared to the max-compute-util policy (which has strict goals to achieve 100% CPU utilization). The second heuristic is the maximum replication factor, which will determine how efficient the cache space utilization will be.

To aid in explaining the scheduling algorithm, we first define several variables:

- $Q$ wait queue
- $T_i$ task at position i in the wait queue; position 0 is the head and position n is the tail
- $E_{set}$ executor sorted set; element existence indicates that the executor is registered and in one of three states: free, busy, or pending
- $I_{map}$ file index hash map; the map key is the file logical name and the value is an executor sorted set of where the file is cached
- $E_{map}$ executor hash map; the map key is the executor name, and the value is a sorted set of logical file names that are cached at the respective executor
- $W$ scheduling window of tasks to consider from the wait queue when making the scheduling decision

The scheduler is separated into two parts, one that sends out a notification, and another that actually decides what task to assign to what executor at the time of work dispatch. The first part of the scheduler takes input a task, and attempts to find the best executor that is free, and notify it that there is work available for pick-up. The pseudo code for this first part is:

```
while (Q !empty)
    for (all files in T0)
        tempSet = Imap(filei)
        for (all executors in tempSet)
            candidates[tempSeti]++
    sort   candidates[] according to values
    for all candidates
        if Eset(candidatei) = freeState
            Mark executor candidatei as pending
            Remove T0 from wait queue and mark as pending
            sendNotificatoin to candidatei to pick up T0
            break
    If no candidate is found in the freeState
        send notification to the next free executor
```

Once an executor receives a notification to pick up a task, assuming it tries to pick up more than one task, the scheduler is invoked again, but this time trying to optimize the lookup given an executor name, rather than a task description. The scheduler then takes the scheduling window size, and starts to build a per task scoring cache hit function. If at any time, a task is found that produces 100% cache hit local rates, the scheduler removes this task from the wait queue and adds it to the list of tasks to dispatch to this executor. This is repeated until the maximum number of tasks were retrieved and prepared to be sent to the executor. If the entire scheduling window is exhausted and no task was found with a cache hit local rate of 100%, the m tasks with the highest cache hit local rates are dispatched.

For the max-compute-util policy, if no tasks were found that would yield any cache hit rates, then the top m tasks are taken from the wait queue and dispatched to the executor. For the max-cache-hit policy, no tasks a returned, signaling that the executor is to return to the free pool of executors. For the good-cache-compute policy, the CPU utilization at the time of scheduling decision will determine which action to take. The CPU utilization is computed by dividing the number of busy nodes with the number of all registered nodes. The pseudo code for the second part is:

```
while (tasksInspected < W)
    fileSet_i = all files in T_i
    cacheHit_i = |intersection fileSet_i and E_map(executor)|
    depending on cacheHit_i and CPU utilization, keep or discard
        keep: remove T_i from Q and add T_i to list to dispatch
        discard: do nothing
    if list of tasks to dispatch is long enough
        break
```

The scheduler's complexity varies with the policy used. For the first-available policy, it is O(1) costs, as it simply takes the first available executor and sends a notification, and dispatches the first task in the queue. The max-cache-hit, max-compute-util, and good-cache-compute policies are more complex with a complexity of $O(|T_i| + replicationFactor + \min(|Q|, W))$. This could equate to 1000s of operations for a single scheduling decision in a worst case, depending on the maximum size of the scheduling window and wait queue length. However, since all data structures used to keep track of executors and files are using hash maps and sorted sets, performing many in-memory operations is quite efficient. Section 5.1 investigated the raw performance of the scheduler under various policies, and we have measured the scheduler's ability to perform 1322 to 1666 scheduling decisions per second for policies (3), (4) and (5) with a maximum window size of 3200.

# 4. ABSTRACT MODEL

We define an abstract model for data-centric task farms as a common parallel pattern that drives the independent computational tasks, taking into consideration the data locality in order to optimize the performance of the analysis of large datasets. The data-centric task farm model is the mirror image of our practical realization in Falkon with its dynamic resource provisioning capabilities and support for data diffusion. Just as Falkon has been used successfully in many domains and applications, we believe our data-centric task farm model generalizes and is applicable to many different domains as well. We claim that the model could help study these concepts of dynamic resource provisioning and data diffusion with greater ease to determine an application end-to-end performance improvements, resource utilization, improved efficiency, and improved scalability. By formally defining this model, we aim for the data diffusion concept to live beyond its practical realization in Falkon. More information on data-centric task farms can be found in a technical report [27].

## 4.1 Base Definitions and Notations

A data-centric task farm has various components that we will formally define in this sub-section.

**Data stores:** Persistent data stores are highly available, scalable, and have large capacity; we assume that data resides on a *set of persistent data stores*, $\Pi$, where $|\Pi| \geq 1$. The *set of transient data stores* $\mathrm{T}$, where $|\mathrm{T}| \geq 0$, are smaller than the persistent data stores and are only capable of storing a fraction of the persistent data stores' data objects. We assume that the transient data stores $\mathrm{T}$ are co-located with compute resources, hence yielding a lower latency data path than the persistent data stores.

**Data Objects:** $\phi(\pi)$ represents the *data objects found in the persistent data store* $\pi$, where $\pi \in \Pi$. Similarly, $\phi(\tau)$ represents a *transient data store's locally cached data objects*. The set of persistent data stores $\Pi$ consists of a *set of all data objects*, $\Delta$. For each *data object* $\delta \in \Delta$, $\beta(\delta)$ denotes the *data object's size* and $\lambda(\delta)$ denotes the *data object's storage location(s).*

**Store Capacity:** For each persistent data store, $\pi \in \Pi$, and transient data store $\tau \in \mathrm{T}$, $\sigma(\pi)$ and $\sigma(\tau)$ denote the persistent and transient *data store's capacity*.

**Compute Speed:** For each transient resource, $\tau \in \mathrm{T}$, $\chi(\tau)$ denotes the *compute speed*.

**Load:** For any data store, we define *load* as the number of concurrent read/write requests; $\omega(\tau)$ and $\omega(\pi)$ denote the load on data stores $\tau \in \mathrm{T}$ and $\pi \in \Pi$.

**Ideal Bandwidth:** For any persistent data store $\pi \in \Pi$, and transient data store $\tau \in \mathrm{T}$, $\nu(\pi)$ and $\nu(\tau)$ denote the *ideal bandwidth for the persistent and transient data store*, respectively. These transient data stores will have limited availability, and the bandwidth is lesser than that of the persistent data stores, $\nu(\tau) < \nu(\pi)$. We assume there are few high capacity persistent data stores and many low capacity transient data stores, such as $\sum_{\tau \in \mathrm{T}} \nu(\tau) \geq \sum_{\pi \in \Pi} \nu(\pi)$, given that $|\mathrm{T}| >> |\Pi|$.

**Available Bandwidth:** For any persistent data store $\pi \in \Pi$, and transient data store $\tau \in \mathrm{T}$, we define *available bandwidth* as a function of ideal bandwidth and load; more formally, $\eta(\nu(\pi), \omega(\pi))$ and $\eta(\nu(\tau), \omega(\tau))$ will denote the available bandwidth for the persistent and transient data store, respectively. The relationship between the ideal and available bandwidth is given by the following formula: $\eta(\nu(\pi), \omega(\pi)) < \nu(\pi)$, for $\omega(\pi) \geq 1$ and $\eta(\nu(\pi), \omega(\pi)) = \nu(\pi)$, for $\omega(\pi) = 0$.

**Copy Time:** For any data object $\delta \in \Delta$ and transient data store $\tau \in \mathrm{T}$, we define the *time to copy a data object* between the object $\delta$ to $\tau$ by the function $\zeta(\delta, \tau) = \begin{cases} \tau_1 \to \tau, \forall \delta \in \phi(\tau_1), \tau_1 \in \mathrm{T} \\ \pi_1 \to \tau, \forall \delta \in \phi(\pi_1), \pi_1 \in \Pi \setminus \mathrm{T} \end{cases}$, where $\tau_1, \pi_1 \to \tau$ denotes the source and destination data stores for the copy operation. In an ideal case, $\tau_1 \to \tau$ can be computed by $\frac{\min[\nu(\tau_1),\nu(\tau)]}{\beta(\delta)}$, where $\nu(\tau_1)$ and $\nu(\tau)$ represent the source and destination *ideal bandwidth*, respectively, and $\beta(\delta)$ represents the data object's size; the same definition applies to copy a data object from $\pi_1 \to \tau$. In reality, this is an oversimplification since copy time $\zeta(\delta,\tau)$ is dependent on other factors such as the load $\omega(\tau)$ on some storage resource, the latency between the source and destination, and the error rates encountered during the transmission. Assuming low error rates and low latency, the copy time is then affected only by the data object's size and the available bandwidth $\eta(\nu(\tau),\omega(\tau))$ as defined above. More formally, $\tau_1 \to \tau$ is defined as $\frac{\min[\eta(\nu(\tau_1),\omega(\tau_1)),\eta(\nu(\tau),\omega(\tau))]}{\beta(\delta)}$.

**Tasks:** Let $\mathrm{K}$ denote the *incoming stream of tasks*. For each *task* $\kappa \in \mathrm{K}$, let $\mu(\kappa)$ denote the *time needed to execute the task* $\kappa$ on the computational resource $\tau \in \mathrm{T}$; let $\theta(\kappa)$ denote the *set of data objects that the task* $\kappa$ *requires*, $\theta(\kappa) \subseteq \Delta$; let $o(\kappa)$ denote the *time to dispatch the task* $\kappa$ *and return a result*.

**Computational Resource State**: If a compute resource $\tau \in \mathrm{T}$ is computing a task, then it is in the *busy* state, denoted by $\tau_b$; otherwise, it is in the free state, $\tau_f$. Let $\mathrm{T}_b$ denote the set of all compute resources in the busy state, and $\mathrm{T}_f$ the set of all compute resources in the free state; these two sets have the following property: $\mathrm{T}_b \cup \mathrm{T}_f = \mathrm{T}$.

## 4.2 The Execution Model

The execution model outlines the policies that control various parts of the execution model and how they relate to the definitions in the previous section. Each incoming task $\kappa \in \mathrm{K}$ is dispatched to a transient resource $\tau \in \mathrm{T}$, selected according to the *dispatch policy*. If a response is not received after a time determined by the *replay policy*, or a failed response is received, the task is re-dispatched according to the *dispatch policy*. A missing data object, $\delta \in \Delta$, that is required by task $\kappa$, $\delta \in \theta(\kappa)$, and does not exist on the transient data store $\tau \in \mathrm{T}$, $\delta \notin \phi(\tau)$, is copied from transient or persistent data stores selected according to the *data fetch policy*. If necessary, existing data at a transient data store $\tau$ are discarded to make room for the new data, according to the *cache eviction policy*. Each computation $\kappa$ is performed on the data objects $\phi(\tau)$ found in a transient data store. We define a *resource acquisition policy* that decides when, how many, and for how long to acquire new transient computational and storage resources for. Similarly, we also define a *resource release policy* that decides when to release some acquired resources.

Each incoming task $\kappa \in \mathrm{K}$ is dispatched to a transient resource $\tau \in \mathrm{T}$, selected according to the *dispatch policy*. We define five dispatch policies: 1) *first-available,* 2) *first-cache-available*, 3) *max-cache-hit*, 4) *max-compute-util,* and 5) *good-cache-compute*. We focus on policy (3) and policy (4) as we already covered the other policies in Section 3.2.

The **max-cache-hit** policy uses information about data location to dispatch each task to executor that yield the highest cache hits. If no preferred executors are free, task dispatch is delayed until a preferred executor becomes available. This policy aims at maximizing the cache hit/miss ratio; a cache hit occurs when a transient compute resource has the needed data on the same transient data store, and a cache miss occurs when the needed data is not the same computational resource's data store. Formally, we define a *cache hit* as follows:

$\forall \delta \in \theta(\kappa), \exists \tau \in \mathrm{T}$, such that $\delta \in \phi(\tau)$.

Similarly, we define a *cache miss* as follows:

$\exists \delta \in \theta(\kappa)$, such that $\forall \tau \in \mathrm{T}, \delta \notin \phi(\tau)$.

Let $C_h(\kappa)$ denote the set of all cache hits, and $C_m(\kappa)$ denote the set of all cache misses for task $\kappa \in \mathrm{K}$, such that $C_h(\kappa) \cup C_m(\kappa) = \theta(\kappa)$. We define the *max-cache-hit* dispatch policy as follows: $\max_{\forall \kappa \in \mathrm{K}}\left(\frac{C_h(\kappa)}{C_m(\kappa)}\right)$.

The **max-compute-util** policy also leverages data location information, but in a different way. It always sends a task to an available executor, but if several workers are available, it selects that one that has the most data needed by the task. This policy aims to maximize computational resource utilization.

We define a *free cache hit* as follows:

$\forall \delta \in \theta(\kappa), \exists \tau \in \mathrm{T}_f$, such that $\delta \in \phi(\tau)$.

Similarly, we define a *free cache miss* as follows:

$\forall \tau \in \mathrm{T}, \delta \notin \phi(\tau)$ or $\exists \tau \in \mathrm{T}_b$, such that $\delta \in \phi(\tau)$.

Let $C_{f,h}(\kappa)$ denote the set of all free cache hits, and $C_{f,m}(\kappa)$ denote the set of all free cache misses for task $\kappa \in \mathrm{K}$, such that $C_{f,h}(\kappa) \subseteq C_h(\kappa)$ and $C_m(\kappa) \subseteq C_{f,m}(\kappa)$ and $C_{f,h}(\kappa) \cup C_{f,m}(\kappa) = \theta(\kappa)$.

We define the *max-compute-util* dispatch policy as follows: $\max_{\forall \kappa \in \mathrm{K}}\left(\frac{C_{f,h}(\kappa)}{C_{f,m}(\kappa)}\right)$.

The **good-cache-compute** policy is the combination of max-compute-util and the max-cache-hit policy, which attempts to strike a good balance between the two policies. This policy was discussed in Section 3.2.

### 4.3 Model Performance and Efficiency

In this section, we investigate when we can achieve good performance with this abstract model for data-centric task farms, and under what assumptions. We define various costs and efficiency related metrics. Furthermore, we explore the relationships between the different parameters in order to optimize efficiency.

**Cost per task:** For simplicity, let us assume initially that each task requires a single data object, $\delta \in \theta(\kappa)$, and that all the data objects $\Delta$ on persistent storage $\Pi$ and transient storage $\mathrm{T}$ are fixed; assume that we use the max-resource-util dispatch policy, then the *cost of the execution of each task* $\kappa \in \mathrm{K}$ dispatched to a transient compute resource $\tau \in \mathrm{T}$ can be characterized as one of the following two costs: 1) *cost if the required data objects are cached* at the corresponding transient storage resource, and 2) *cost if the required data objects are not cached* and must be retrieved from another transient or persistent data store. In the first case, we define the cost of the execution of a task to be the time to dispatch the task plus the time to execute the task plus the time to return the result. For the second cost function in which the data objects do not exist in the transient data store, we also incur an additional cost to copy the needed data object from either a persistent or a transient data store. More formally, we define the *cost per task* $\chi(\kappa)$ as:

$$\chi(\kappa) = \begin{cases} o(\kappa) + \mu(\kappa), & \delta \in \phi(\tau) \\ o(\kappa) + \mu(\kappa) + \zeta(\delta,\tau), & \delta \notin \phi(\tau) \end{cases}$$

**Average Task Execution Time:** We define the *average task execution time,* $\mathrm{B}$, as the summation of all the task execution times divided by the number of tasks; more formally, we have $\mathrm{B} = \frac{1}{|\mathrm{K}|}\sum_{k \in \mathrm{K}} \mu(\kappa)$.

**Computational Intensity:** Let $\mathrm{A}$ denote the *arrival rate of tasks*; we define the *computational intensity,* $\mathrm{I}$, as follows: $\mathrm{I} = \mathrm{B} * \mathrm{A}$. If $\mathrm{I} = 1$, then all nodes are fully utilized; if $\mathrm{I} > 1$, tasks are arriving faster than they can be executed; finally, if $\mathrm{I} < 1$, it indicates idle nodes.

**Workload Execution Time**: We define the *workload execution time,* $V$, of our system as $V = \max\left(\frac{B}{|\mathrm{T}|}, \frac{1}{\mathrm{A}}\right) * |\mathrm{K}|$.

**Workload Execution Time with Overhead**: In general, the total execution time for a task $\kappa \in \mathrm{K}$ includes overheads, which reduced efficiency by a factor of $\frac{\mu(\kappa)}{\chi(\kappa)}$. We define the *workload execution time with overhead,* $W$, of our system as $W = \max\left(\frac{\mathrm{Y}}{|\mathrm{T}|}, \frac{1}{\mathrm{A}}\right) * |\mathrm{K}|$, where $Y$ is the *average task execution time including overheads* defined as

$$Y = \begin{cases} \frac{1}{|\mathrm{K}|}\sum_{\kappa \in \mathrm{K}} [\mu(\kappa) + o(\kappa)], & \delta \in \phi(\tau), \delta \in \Omega \\ \frac{1}{|\mathrm{K}|}\sum_{\kappa \in \mathrm{K}} [\mu(\kappa) + o(\kappa) + \zeta(\delta,\tau)], & \delta \notin \phi(\tau), \delta \in \Omega \end{cases}.$$

**Efficiency:** We define the *efficiency,* $\mathrm{E}$, of a particular workload as $\mathrm{E} = \frac{V}{W}$. The expanded version of efficiency is $\mathrm{E} = \frac{\max\left(\frac{B}{|\mathrm{T}|}, \frac{1}{\mathrm{A}}\right) * |\mathrm{K}|}{\max\left(\frac{\mathrm{Y}}{|\mathrm{T}|}, \frac{1}{\mathrm{A}}\right) * |\mathrm{K}|}$, which can be reduced to

$$E = \begin{cases} 1, & \frac{Y}{|T|} \le \frac{1}{A} \\ \max\left(\frac{B}{Y}, \frac{|\mathrm{T}|}{\mathrm{A} * Y}\right), & \frac{Y}{|T|} > \frac{1}{A} \end{cases}.$$

We claim that for the caching mechanisms to be effective, the *aggregate capacity of our transient storage resources* $\mathrm{T}$ *must be greater than our workload's working set,* $\Omega$, *size*; formally $\sum_{\tau \in \mathrm{T}} \sigma(\tau) \ge |\Omega|$.

We claim that we can obtain $\mathrm{E} > 0.5$ if $\mu(\kappa) > o(\kappa) + \zeta(\delta,\tau)$, where $\mu(\kappa)$, $o(\kappa)$, $\zeta(\delta,\tau)$ are the time to execute and dispatch the task $\kappa \in \mathrm{K}$, and copy the object $\delta$ to $\tau \in \mathrm{T}$, respectively.

**Speedup:** We define the *speedup,* $S$, of a particular workload as $S = E * |T|$.

**Optimizing Efficiency:** Having defined both efficiency and speedup, it is possible to maximize for either one. We *optimize efficiency* by finding the smallest number of *transient compute/storage resources* $|\mathrm{T}|$ while maximizing speedup*efficiency.

### 4.4 Model Validation

We perform a preliminary validation of our abstract model with results from a real large-scale astronomy application [5, 6]. We found the model to be relatively accurate for a wide range of empirical results we obtained from an astronomy application. For 92 experiments from [31], the model error is good (5% average and 5% median) with a standard deviation of 5%, and a worst case model error of 29%.

Figure 2 shows the details of the model error under the various experiments. These experiments were from an astronomy application which had a working set of 558,500 files (1.1TB compressed and 3.35TB uncompressed). From this working set, various workloads were defined that had certain data locality characteristics, varying from the lowest locality of 1 (i.e., 1-1 mapping between objects and files) to the highest locality of 30 (i.e., each file contained 30 objects).

Figure 2 (left) shows the model error for experiments that varied the number of CPUs from 2 to 128 with locality of 1, 1.38, and 30. Note that each model error point represents a workload that spanned 111K, 154K, and 23K tasks for data locality 1, 1.38, and 30 respectively. The second set of results (Figure 2 - right) fixed the number of CPUs at 128, and varied the data locality from 1 to 30. The results show a larger model error with an average of 8% and a standard deviation of 5%. We attribute the model errors to contention in the shared file system and network resources that are only captured simplistically in the current model.

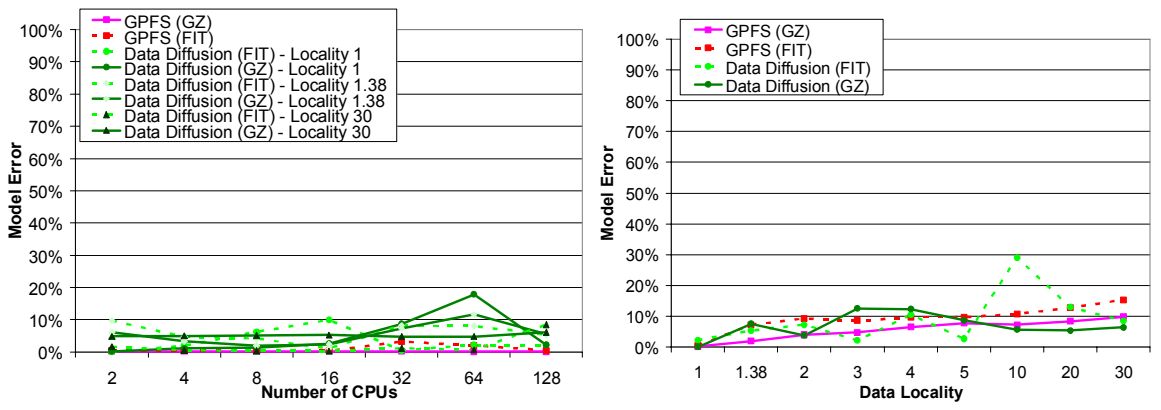

***Figure 2: Model error for varying # of CPU and data-locality***

The second set of results (Figure 2 - right) fixed the number of CPUs at 128, and varied the data locality from 1 to 30. The results here show a larger model error with an average of 8% and a standard deviation of 5%. We attribute the model errors to contention in the shared file system and network resources that are only captured simplistically in the current model.

We also plan to do a thorough validation of the model through discrete-event simulations that will allow us to investigate a wider parameter space than we could in a real world implementation. Through simulations, we also hope to measure application performance in a more dynamic set of variables that aren't bound to single static values, but could be complex functions inspired from real world systems and applications. The simulations will specifically attempt to model a Grid environment comprising of computational resources, storage resources, batch schedulers, various communication technologies, various types of applications, and workload models. We will perform careful and extensive empirical performance evaluations in order to create correct and accurate input models to the simulator; the input models include 1) Communication costs, 2) Data management costs, 3) Task scheduling costs, 4) Storage access costs, and 5) Workload models. The outputs from the simulations over the entire considered parameter space will form the datasets that will be used to statistically validate the model using $R^2$ ***statistic*** and *graphical residual analysis* [**33**]

# 5. EMPIRICAL EVALUATION

We conducted several experiments to understand the performance and overhead of the data-aware scheduler, as well as to see the effect of dynamic resource provisioning and data diffusion. The experiments ran on the ANL/UC TeraGrid [18, 22] site using 64 nodes. The Falkon service ran on gto.ci.uchicago.edu (8-core Xeon @ 2.33GHz per core, 2GB RAM, Java 1.6) with 2 ms latency to the executor nodes.

We performed a wide range of experiments that covered various scheduling policies and settings. In all experiments, the data is originally located on a GPFS [8] shared file system with sub 1ms latency. We investigated the performance of 4 policies: 1) first-available, 2) max-cache-hit, 3) max-compute-util, and 4) good-cache-compute. In studying the effects of dynamic resource provisioning on data diffusion, we also investigated the effects of the cache size, by varying the per node cache size from 1GB, 1.5GB, 2GB, to 4GB.

## 5.1 Scheduler

In order to understand the performance of the data-aware scheduler, we developed several micro-benchmarks to test scheduler performance. We used the first-available policy that performed no I/O as the baseline scheduler, and tested the various scheduling policies. We measured overall achieved throughput in terms of scheduling decisions per second and the breakdown of where time was spent inside the Falkon service. We conducted our experiments using 32 nodes; our workload consisted of 250K tasks, where each task accessed a random file (uniform distribution) from a dataset of 10K files of 1B in size each. We use files of 1 byte to measure the scheduling time and cache hit rates with minimal impact from the actual I/O performance of persistent storage and local disk. We compare the first-available policy using no I/O (sleep 0), first-available policy using GPFS, max-compute-util policy, max-cache-hit policy, and good-cache-compute policy. The scheduling window size was set to 100X the number of nodes, or 3200. We also used 0.8 as the CPU utilization threshold in the good-cache-compute policy to determine when to switch between the max-cache-hit and max-compute-util policies.

Figure 3 shows the scheduler performance under different scheduling policies. We see the throughput in terms of scheduling decisions per second range between 2981/sec (for first-available without I/O) to as low as 1322/sec (for max-cache-hit).

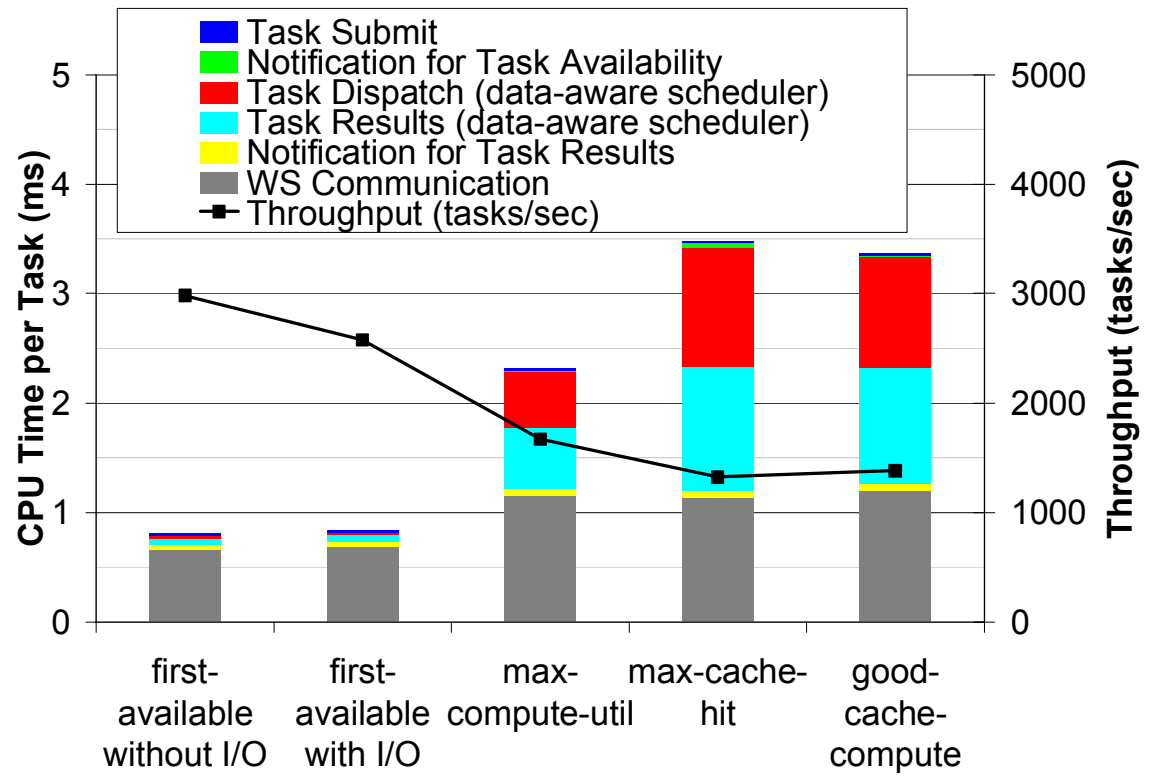

***Figure 3: Data-aware scheduler performance and code profiling for the various scheduling policies***

It is worth pointing out that for the first-available policy, the cost of communication is significantly larger than the rest of the costs combined, including scheduling. The scheduling is quite inexpensive for this policy as it simply load balances across all workers. However, we see that with the 3 data-aware policies, the scheduling costs (red and light blue areas) are more significant.

## 5.2 Provisioning

The key contribution of this paper is the study of dynamic resource provisioning in the context of data diffusion, and how it performs for data intensive workloads. In choosing our workload, we set the I/O to compute ratio large (10MB of I/O to 10ms of compute). The dataset consists of 10K files, with 10MB per file. Each task reads one file chosen at random from the dataset, and computes for 10ms. The workload had an initial arrival rate of 1 task/sec, a multiplicative increasing function by 1.3, 60 seconds between increase intervals, and a maximum arrival rate of 1000 tasks/sec. The increasing function is $A_i = \min\left[ceiling(A_{i-1} * 1.3), 1000\right], 0 \le i < 24$, which varies arrival rate A from 1 to 1000 in 24 distinct intervals making up 250K tasks and spanning 1415 seconds to complete. This workload is both data-intensive and has good locality of reference, a good candidate to measure the impact of data diffusion and resource provisioning.

Note that we needed a high I/O to compute ratio due to the small testbed we used (64 nodes). For example, if we were to set the ratio to a more balanced value, 1MB I/O and 1 second compute, having 64 dual processor nodes would achieve at most 128MB/s (1Gb/s). GPFS can sustain 4Gb/s+ of read rates, which would have meant that on our testbed, GPFS performance would have been sufficient. When we get access to a larger testbed and we scale up the experiments to 100s or 1000s of nodes, we'll be able to explore more balanced I/O to compute ratios while still requiring more throughput than shared file systems can deliver.

### *5.2.1 Cache Size Effects on Data Diffusion*

We begin the data diffusion results with the summary view of several experiments showing the effects of the cache size on the performance of executing the workload. We also show the baseline execution of the first-available policy, which does not use data diffusion, and simply load balances across the nodes tasks that work directly on the shared file system.

Several measured or computed metrics are relevant in understanding the following set of graphs. These include ideal throughput, throughput, number of nodes, wait queue length, cache hit local/global %, and cache miss %. They are defined as follows:

*Ideal Throughput (Gb/s)*: throughput needed to satisfy arrival rate; A*fileSize per some unit time

*Throughput (Gb/s)*: measured aggregate throughput; successfulTasks*fileSize per some unit time

*Number of Nodes (N)*: number of registered nodes; i.e., the maximum number of nodes that can execute tasks at once (2 per node, 1 per CPU)

*Wait Queue Length*: number of tasks in the wait queue

*Cache Hit Global % ($HR_C$)*: global cache hits are file accesses that required the file to be transferred from another worker cache; $HR_C = H_C/(H_L + H_C + H_S)$

*Cache Hit Local % ($HR_L$)*: local cache hits are file accesses that can be served entirely from local cache (i.e. local disk); $HR_L = H_L/(H_L + H_C + H_S)$

*Cache Miss % ($HR_S$)*: cache misses are file accesses that are not found in any worker cache, and have to be served from the shared file system (i.e. GPFS); $HR_S = H_S/(H_L + H_C + H_S)$

Figure 4 shows the baseline experiment (first-available policy). This experiment ran the workload of 250K tasks, where each task worked directly on the shared file system (LAN GPFS), the common practice in many scientific applications. We used dynamic resource provisioning which allocated resources on demand based on load. The load metric was the wait queue length (denoted by the thin pink line); note that a short wait queue length is desirable, indicating that the resources are able to process tasks as they arrive. The black monotonically increasing line denotes the number of nodes provisioned. Finally, we have two throughput metrics we show, one is the ideal (light blue line) and the other is the measured aggregate throughput (dark blue line). Recall that the ideal throughout is the throughput needed to satisfy the arrival rate.

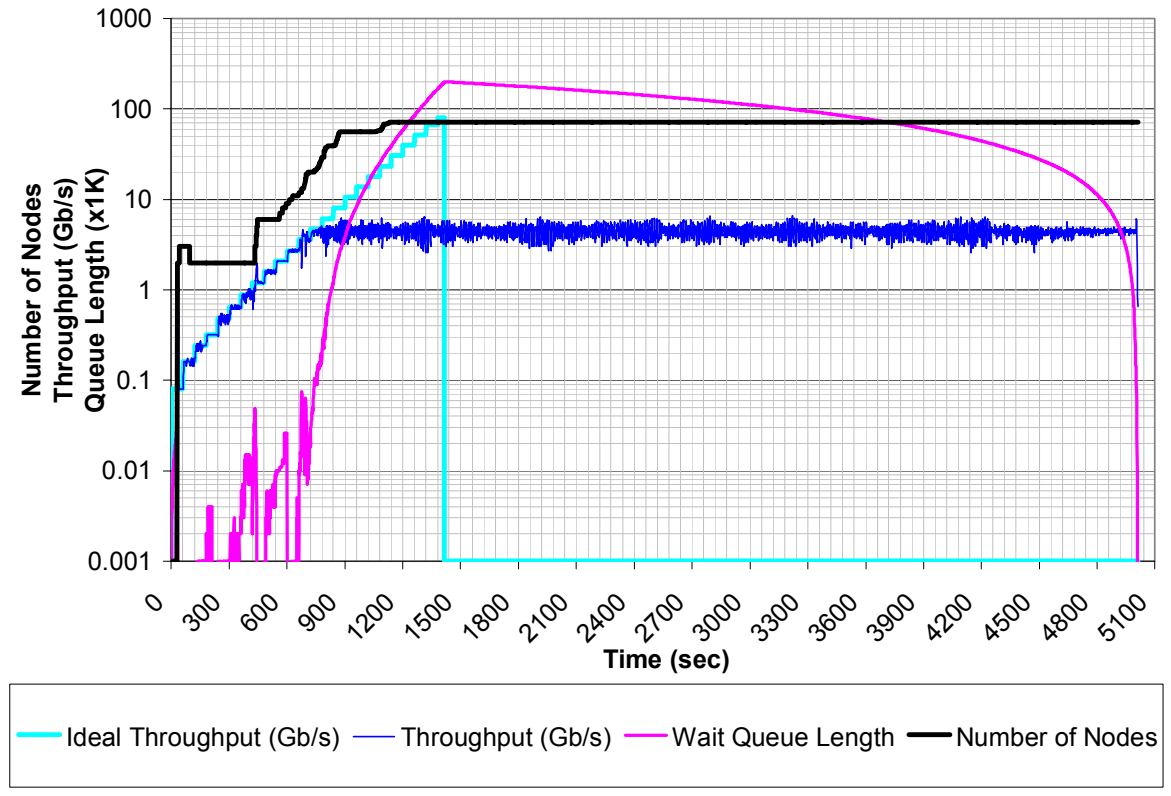

***Figure 4: Summary view of 250K tasks executed via the first-available policy directly on GPFS using dynamic resource provisioning***

Aggregate throughput matches the ideal throughput for arrival rates ranging between 1 and 59 tasks/sec, but the throughput remains flat at an average of 4.4Gb/s for greater arrival rates. At the transition point when the arrival rate increased beyond 59, the wait queue length also started growing beyond the relatively small values, to the eventual length of 198K tasks. The workload execution time was 5011 seconds, which yielded 28% efficiency (with ideal time being 1415 seconds).

Figure 5-8 (similar to Figure 4) summarizes results for data diffusion with varying cache sizes per node (1GB, 1.5GB, 2GB, and 4GB) using the good-cache-compute policy; recall that this policy is a combination between the max-cache-hit and max-compute-util policy, which attempts to optimize the cache hit performance as long as processor utilization is high (80% in our case). The dataset originally resided on the GPFS shared file system, and was diffused to local disk caches with every cache miss (the red area in the graphs); cache hit global (file accesses from remote worker caches) rates are shown in yellow, while the cache hit local (file accesses satisfied from the local disk) rates are shown in green.

Figure 5 is an interesting use case as it shows the performance of data diffusion when the working set does not fit in cache. In our case, the working set was 100GB, but the aggregate cache size was 64GB as we had 64 nodes at the peak of the experiment. Notice that throughput keeps up with the ideal throughput for a little longer than the first-available policy, up to 101 tasks/sec arrival rates. At this point, the throughput stabilizes at an average of 5.2Gb/s until 800 seconds later when the cache hit rates increase due to the working set caching reaching a steady state, when the throughput at an average of 6.9Gb/s. The overall cache hit rate was 31%, which in the end resulted in a 57% higher throughput than what the first-available policy was able to achieve using GPFS directly. Also, note that the workload execution time is reduced to 3762 seconds, down from 5011 seconds for the first-available policy; the efficiency when compared to the ideal case is 38%.

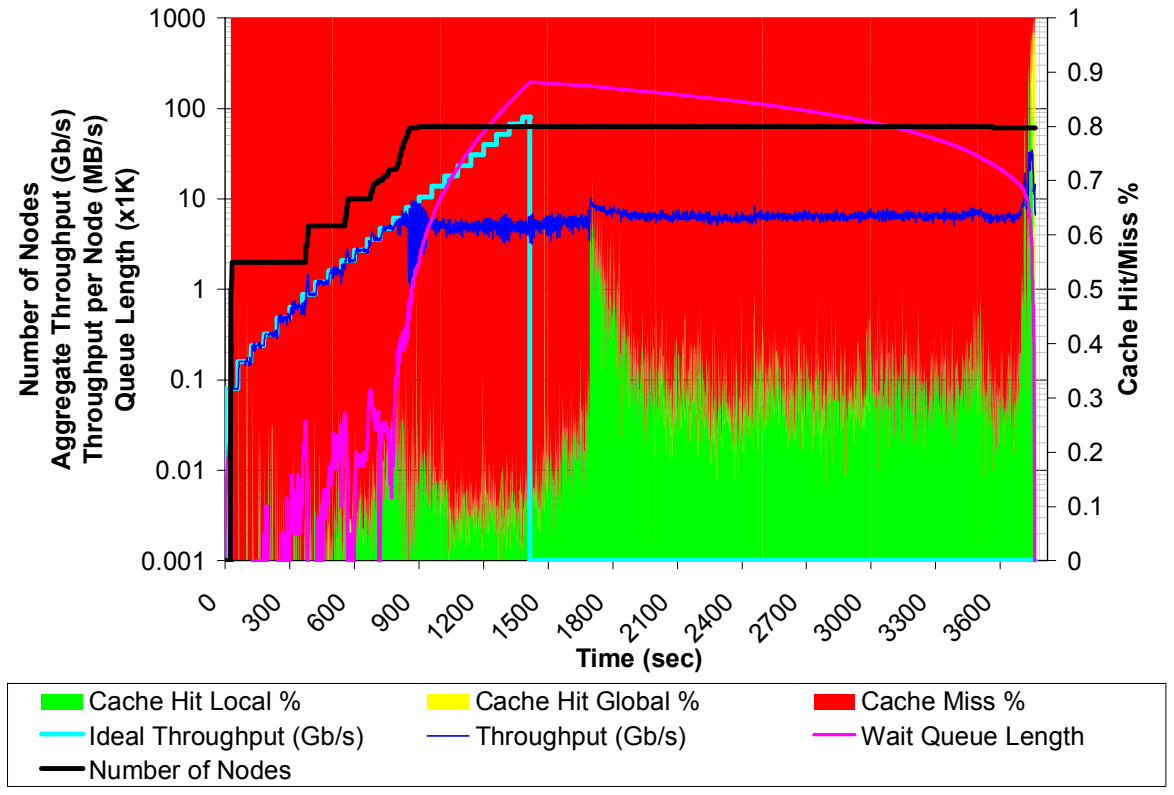

***Figure 5: Summary view of 250K tasks executed using data diffusion and good-cache-compute policy with 1GB caches per node and dynamic resource provisioning***

Figure 6 increases the per node cache size from 1Gb to 1.5GB, which increases the aggregate cache size to 96GB, almost enough to hold the entire working set of 100GB.

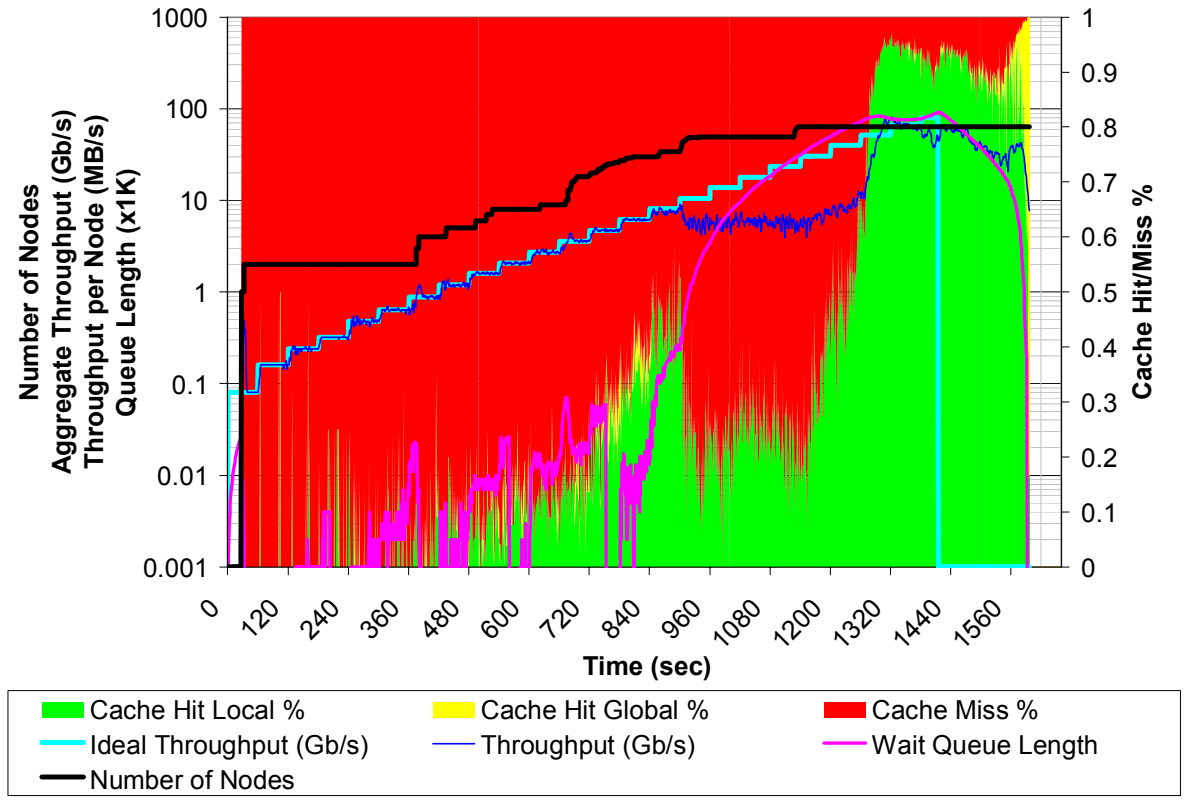

***Figure 6: Summary view of 250K tasks executed using data diffusion and good-cache-compute policy with 1.5GB caches per node and dynamic resource provisioning***

Notice that the throughput hangs on further to the ideal throughput, up to 132 tasks/sec when the throughput increase stops and stabilizes at an average of 6.3Gb/s. Within 350 seconds of this stabilization, the cache hit performance increased significantly from 25% cache hit rates to over 90% cache hit rates; this increase in cache hit rates also results in the throughput increase up to an average of 45.6Gb/s for the remainder of the experiment. Overall, it achieved 78% cache hit rates, 1% cache hit rates to remote caches, and 21% cache miss rates. Overall, the workload execution time was reduced drastically from the 1GB per node cache size, down to 1596 seconds; this yields a 89% efficiency when compared to the ideal case.

Figure 7 increases the cache size further to 2GB per node, for a total of 128GB which was finally large enough to hold the entire working set of 100GB. We see the throughput is able to hold onto the ideal throughput quite well for the entire experiment. The great performance is attributed to the ability to cache the entire working set, and schedule tasks to the nodes that had the data cached approaching with cache hit rates of 98%. Its also interesting to note that the queue length never grew beyond 7K tasks long, which was quite a feat given that the other experiments so far (first-available policy, and good-cache-compute with 1GB and 1.5GB caches) all ended up with queues in the 91K to 200K tasks long. With an execution time of 1436 seconds, the efficiency was 99% of the ideal case.

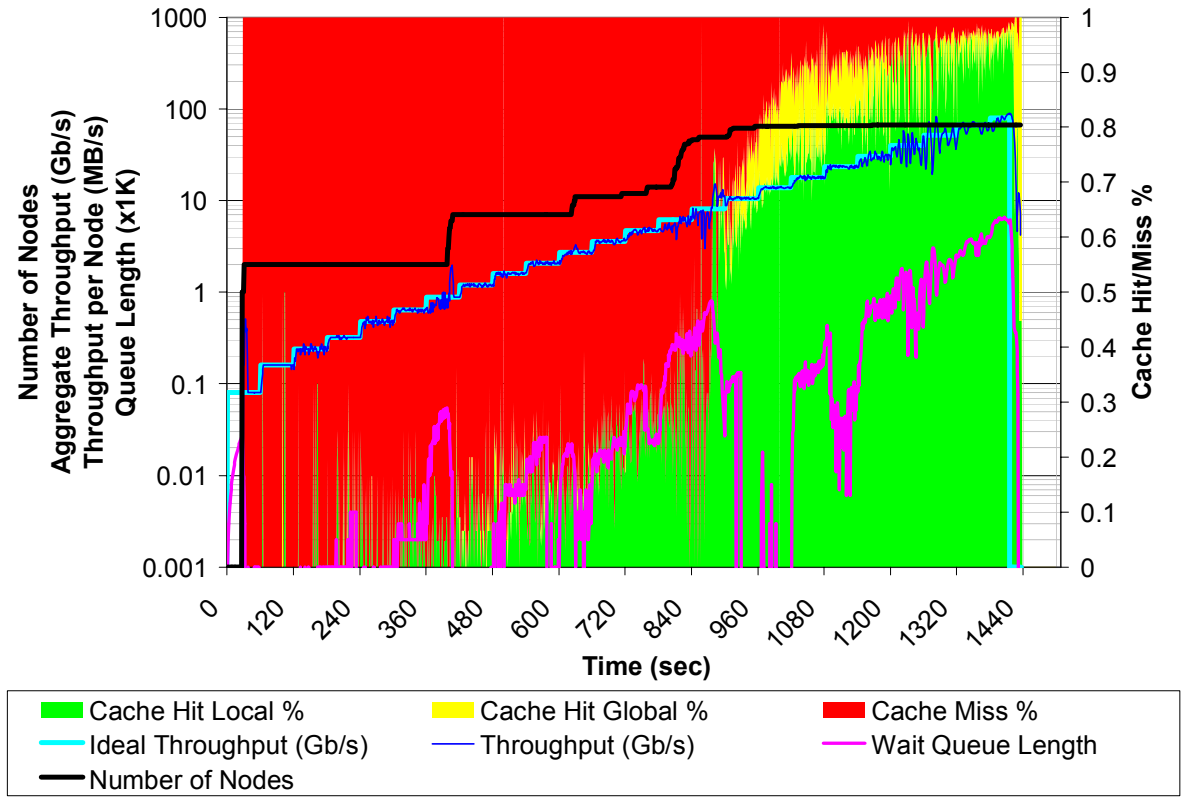

***Figure 7: Summary view of 250K tasks executed using data diffusion and good-cache-compute policy with 2GB caches per node and dynamic resource provisioning***

Investigating if it helps to increase the cache size further to 4GB per node, we conduct the experiment whose results are found in Figure 8. We see no significant improvement in performance.

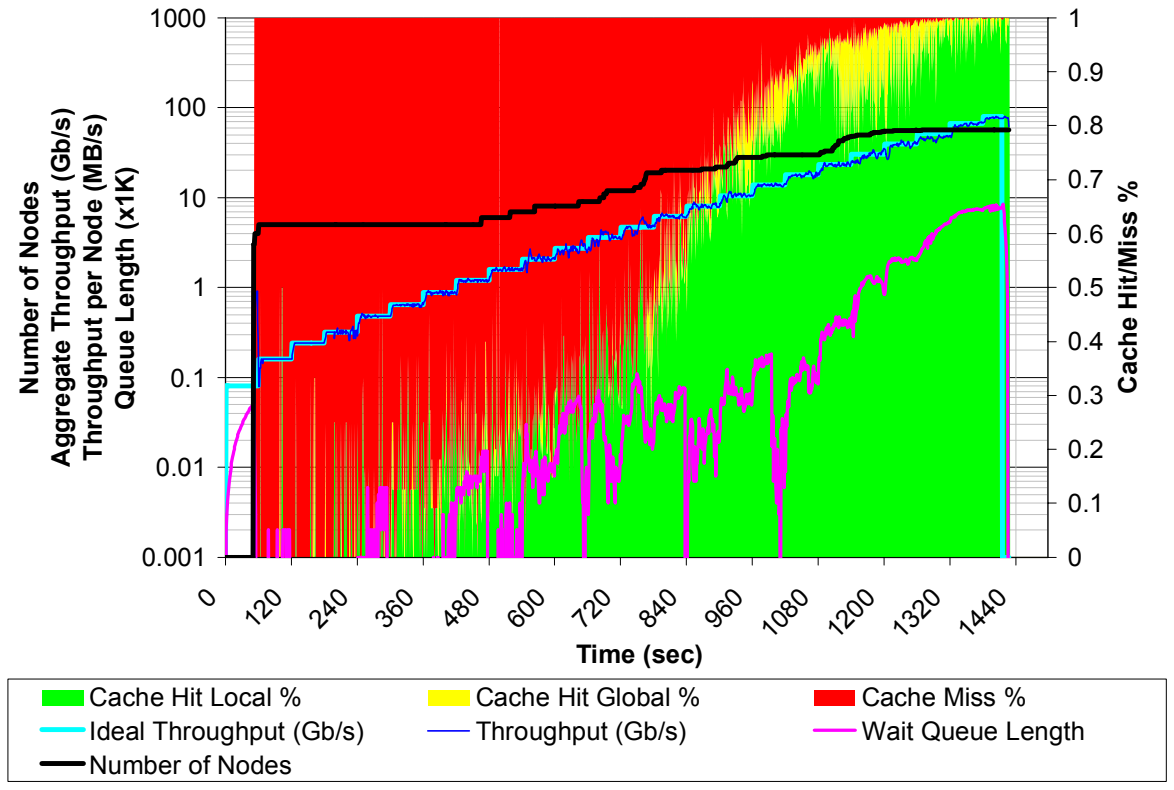

***Figure 8: Summary view of 250K tasks executed using data diffusion and good-cache-compute policy with 4GB caches per node and dynamic resource provisioning***

The execution time is reduced slightly to 1427 seconds (99% efficient), and the overall cache hit rates are improved to 88% cache hit rates, 6% remote cache hits, and 6% cache misses. In order to show the need for the good-cache-compute policy (the previous results from Figure 5 through Figure 8), which is a combination of the max-cache-hit and max-compute-util policy, it is interesting to show the performance for each of these two policies. We fixed the cache size per node at 4GB in order to give both policies ample opportunity for good performance.

Figure 9 shows the performance of the max-cache-hit policy which always schedules tasks according to where the data is cached, even if it has to wait for some node to become available, leaving some nodes processors idle. Notice a new metric measured (dotted thin black line), the CPU utilization, which shows clear poor CPU utilization that decreases with time as the scheduler has difficulty scheduling tasks to busy nodes; the average CPU utilization for the entire experiment was 43%.

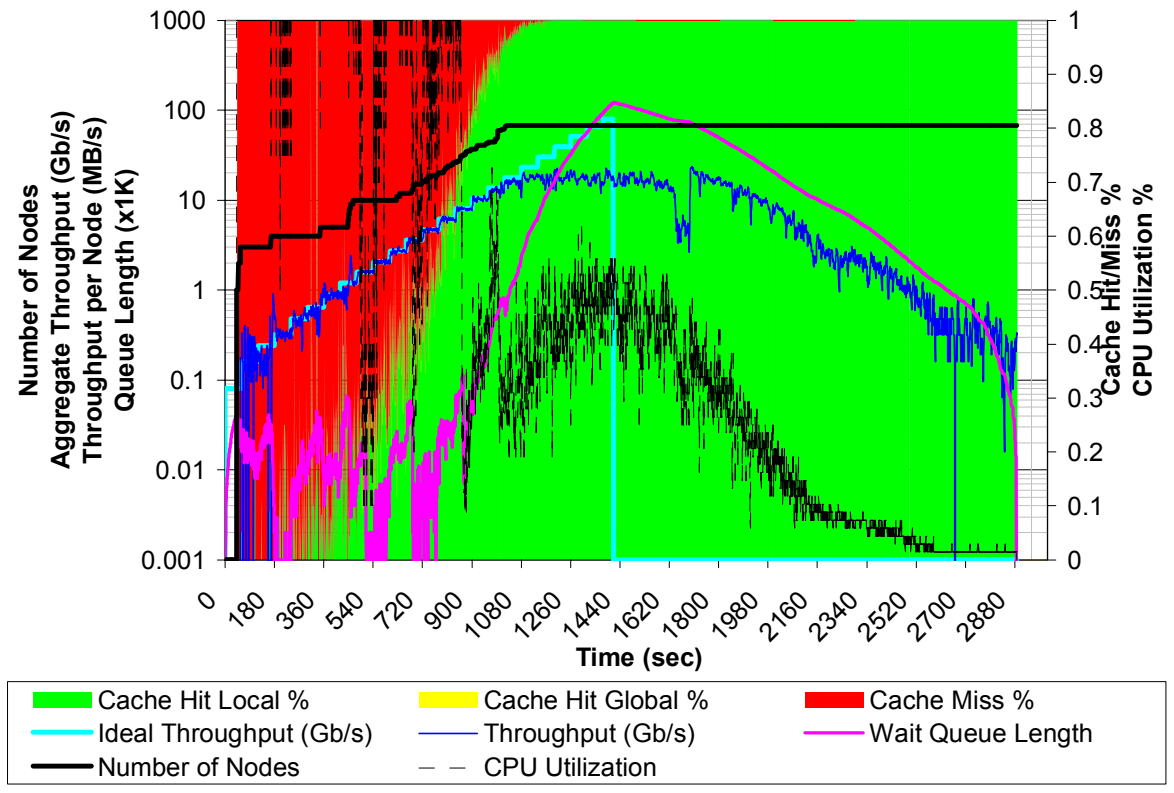

***Figure 9: Summary view of 250K tasks executed using data diffusion and max-cache-hit policy with 4GB caches per node and dynamic resource provisioning***

Its interesting to compare with the good-cache-compute policy which achieved good cache hit rates (88%) at the cost of only 4.5% idle CPUs. However, it's important to point out that the goal of the policy to maximize the cache hit rates was met, as it achieved 94.5% cache hit rates and 5.5% cache miss rates. The workload execution time was a bit disappointing (but not surprising base on the CPU utilization) with 2888 seconds (49% of ideal).

Our final experiment looked at the max-compute-util policy, which attempted to maximize the CPU utilization at the expense of data movement. We see the workload execution time is improved (compared to max-cache-hit) down to 2037 seconds (69% efficient), but it is still far from the good-cache-compute policy that achieved 1436 seconds. The major difference here is that the there are significantly more cache hits to remote caches as tasks got scheduled to nodes that didn't have the needed cached data due to being busy with other work. We were able to sustain high efficiency with arrival rates up to 380 tasks/sec, with an average throughput for the steady part of the experiment of 14.5 Gb/s. It is interesting to see the cache hit local performance at time 1800~2000 second range spiked from 60% to 98%, which results in a spike in throughout from 14Gb/s to 40Gb/s. Although we maintained 100% CPU utilization, due to the extra costs of moving data from remote executors, the performance was worse than the good-cache-compute policy when 4.5% of the CPUs were left idle.

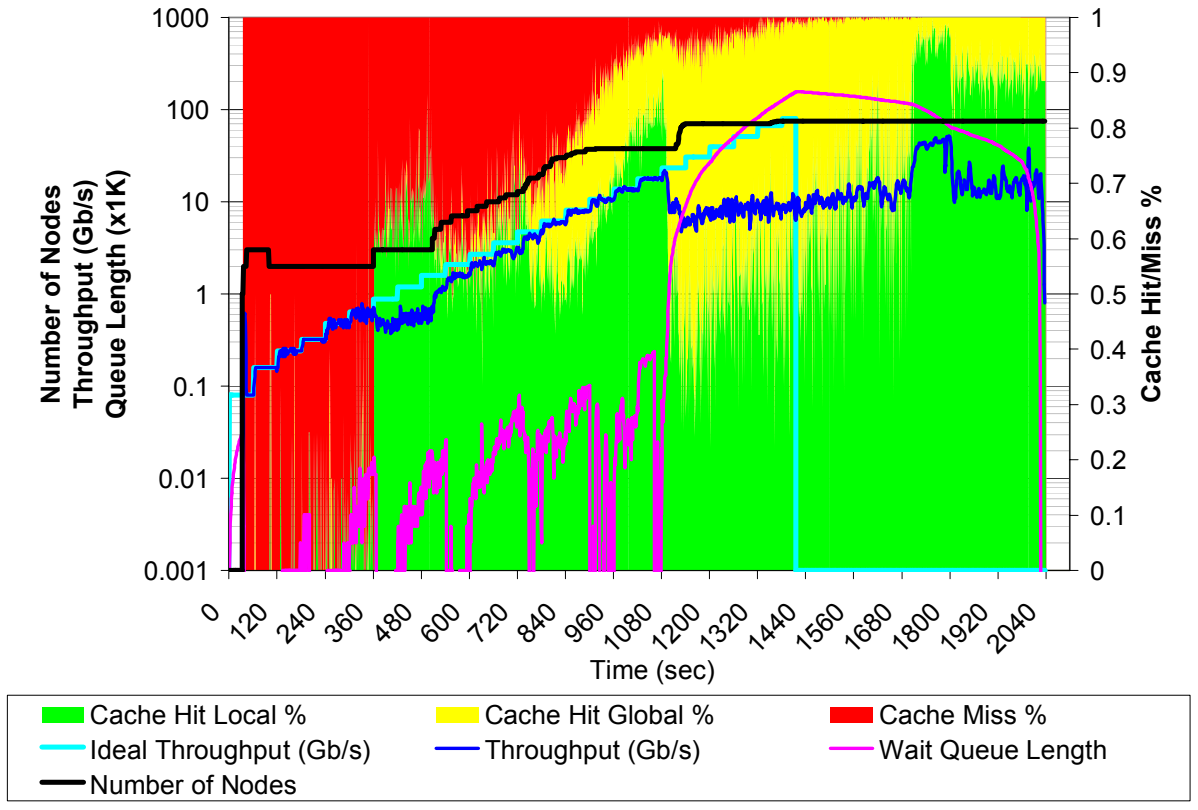

***Figure 10: Summary view of 250K tasks executed in a LAN using data diffusion and max-compute-util policy with 4GB caches per node and dynamic resource provisioning***

### *5.2.2 Cache Performance*

Figure 11 shows cache performance over six experiments involving data diffusion, the ideal case, and the first-available policy which does not cache any data.

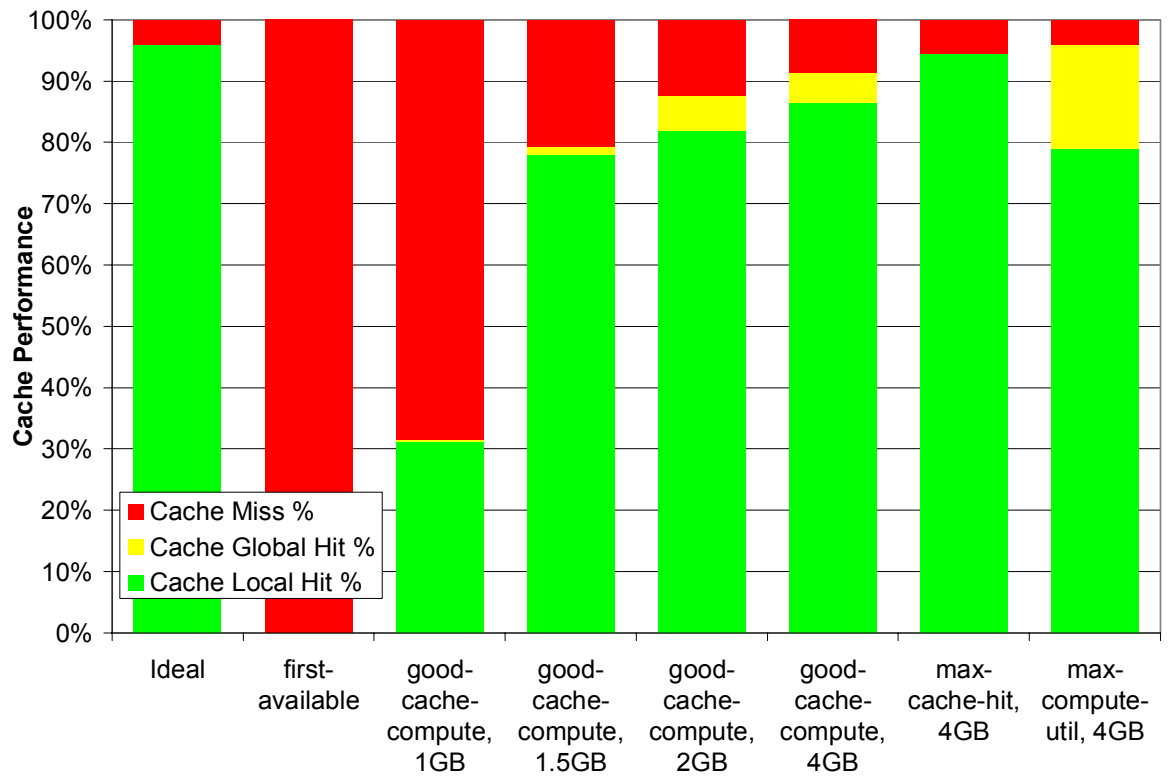

***Figure 11: Cache performance for both LAN and WAN***

We see a clear separation in the cache miss rates (red) for the cases where the working set fit in cache (1.5GB and greater), and the case where it did not (1GB). For the 1GB case, the cache miss rate was 70%, which is to be expected considering only 70% of the working set fit in cache at most, and cache thrashing was hampering the scheduler's ability to achieve better cache miss rates. The other extreme, the 4GB cache size cases, all achieved near perfect cache miss rates of 4%~5.5%.

### *5.2.3 Throughput*

Figure 12 compares the throughputs (broken down into three categories, local cache, remote cache, and GPFS) of all 7 experiments presented in Figure 4 through Figure 10, and how they compare to the ideal case. The first-available policy had the lowest average throughput of 4Gb/s, compared to between 5.3Gb/s and 13.9Gb/s for data diffusion, and 14.1Gb/s for the ideal case. In addition to having much higher average throughputs, data diffusion experiments also achieved significantly higher peak throughputs (the black bar): as high as 100Gb/s as opposed to 6Gb/s for the first-available policy.

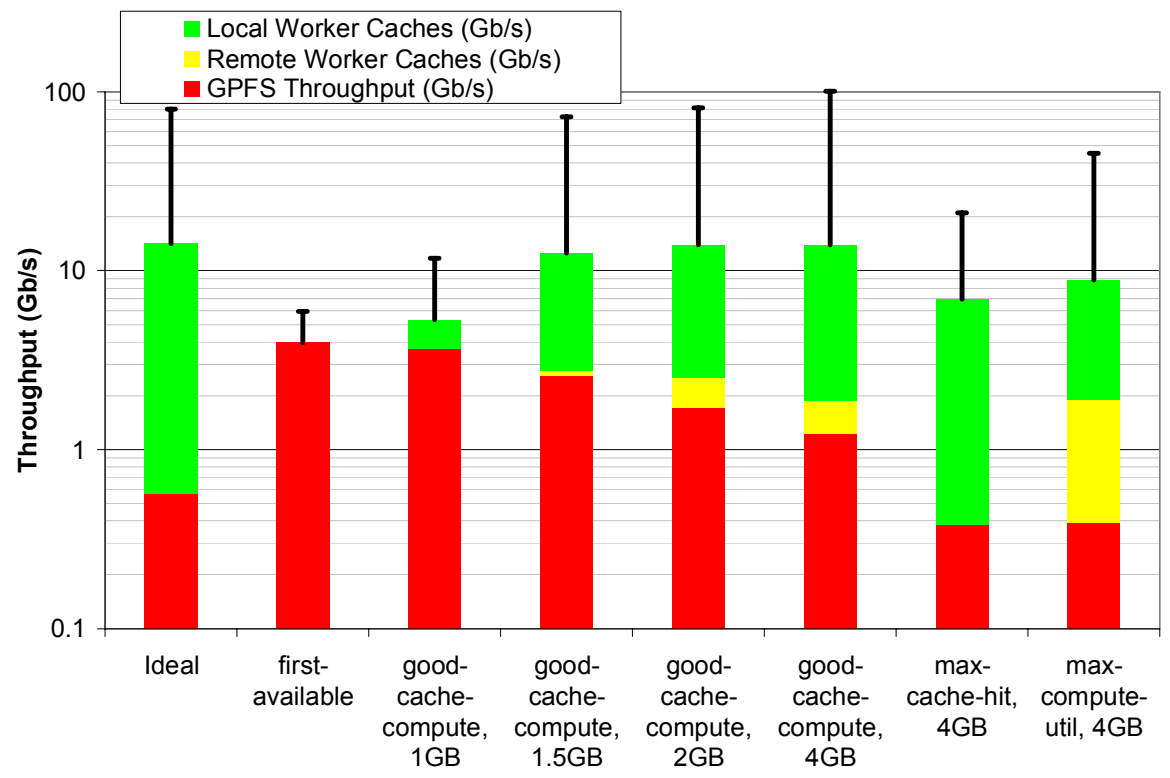

***Figure 12: Average and peak (99 percentile) throughput for both LAN and WAN***

Note also that GPFS file system load (the red portion of the bars) is significantly lower with data diffusion than for the GPFS-only experiments; in the worst case, with 1GB caches where the working set did not fit in cache, the load on GPFS is still high with 3.6Gb/s due to all the cache misses, while GPFS-only tests had 4Gb/s load. However, as the cache sizes increased and the working set fit in cache, the load on GPFS reached as low as 0.4Gb/s. Even the network load due to remote cache access was considerably low, with the highest values of 1.5Gb/s for the max-compute-util policy. All other experiments had less than 1Gb/s network load due to remote cache access.

### *5.2.4 Performance Index and Speedup*

The performance index attempts to capture the speedup per CPU time achieved:

> *Speedup (SP)*: SP measures the improved workload execution time (WET) for the data diffusion (DD) approach as compared to the baseline shared file system (GPFS) approach suing the first-available policy; $SP = WET_{GPFS}/WET_{DD}$
>
> *CPU Time ($CPU_T$)*: the amount of CPU time used
>
> *Performance Index (PI)*: attempts to capture the performance per CPU hour achieved; $PI=SP/CPU_T$, and is normalized for values between 0 and 1 for easier comparisons

Figure 13 shows PI and speedup data. Notice that while both the good-cache-compute with 2GB and 4GB caches achieves the highest speedup of 3.5X, the 4GB case achieves a higher performance index of 1 as opposed to 0.7 for the 2GB case. This is due to the fact that fewer resources were used throughput the 4GB experiment, 17 CPU hours instead of 24 CPU hours for the 2GB case. This reduction in resource usage was due to the larger caches, which in turn allowed the system to perform better with fewer resources for longer durations, and hence the wait queue didn't grow as fast, which resulted in less aggressive resource allocation.

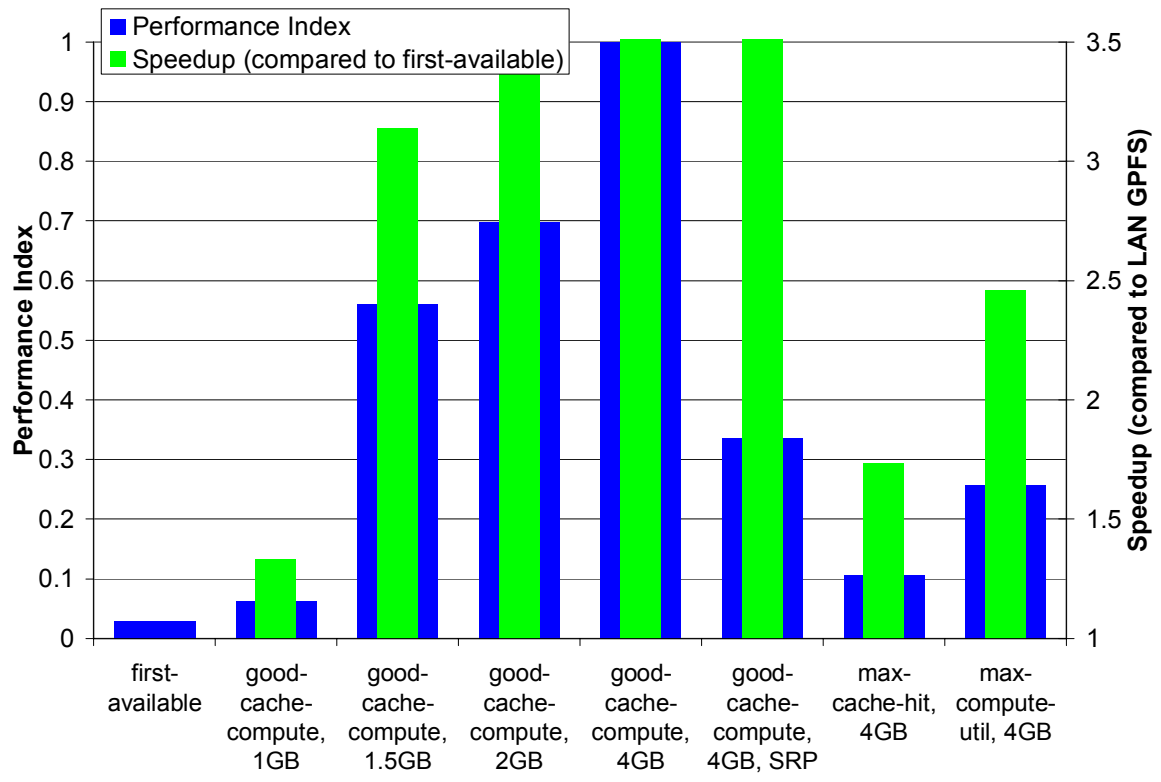

***Figure 13: PI and speedup data for both LAN and WAN***

For comparisons, we also ran the best performing experiment (good-cache-compute with 4GB caches) without dynamic resource provisioning, in which case we allocated 64 nodes ahead of time outside the experiment measurement and maintained 64 nodes throughout the experiment. Notice the speedup is identical to that of using dynamic resource provisioning, we see the performance index is quite low (0.33) due to the additional CPU time that was consumed (46 CPU hours as opposed to 17 CPU hours for the dynamic resource provisioning case). Finally, notice the performance index of the first-available policy which uses GPFS solely; although the speedup gains with data diffusion compared to the first-available policy are relatively modest (1.3X to 3.5X), the performance index of data diffusion is much more, from at least 2X to as high as 34X.

### *5.2.5 Slowdown*

Speedup compares data diffusion to the base case of the LAN GPFS, but does not tell us how well data diffusion performed in relation to the ideal case. Recall that the ideal case is computed from the arrival rate of tasks, assuming zero communication costs and infinite resources to handle tasks in parallel; in our case, the ideal workload execution time is 1415 seconds. Figure 14 shows the slowdown for the LAN experiments as a function of arrival rates. *Slowdown (SL)* measures the factor by which the workload execution times are slower than the ideal workload execution time; the ideal workload execution time assumes infinite resources and 0 cost communication, and is computed from the arrival rate function; $SL=WET_{policy}/WET_{ideal}$; in our case, $WET_{ideal}$ is 1415 seconds.

These results in Figure 14 clearly show the arrival rates that could be handled by each approach, showing the first-available policy (the GPFS only case) to saturate the earliest at 59 tasks/sec denoted by the rising red line. It is evident that larger cache sizes allowed the saturation rates to be higher (essentially perfect for some cases, such as the good-cache-compute with 4GB caches). It interesting to point out the good-cache-compute policy with 1.5GB caches slowdown increase relatively early (similar to the 1GB case), but then towards the end of the experiment the slowdown is reduced from almost 5X back down to an almost ideal 1X. This sudden improvement in performance is attributed to a critical part of the working set being cached and the cache hit rates increasing significantly. Also, note the odd slowdown (as high as 2X) of the 4GB cache DRP case at arrival rates 11, 15, and 20; this slowdown matches up to the drop in throughput between time 360 and 480 seconds in Figure 10 (the detailed summary view of this experiment), which in turn occurred when an additional resource was allocated.

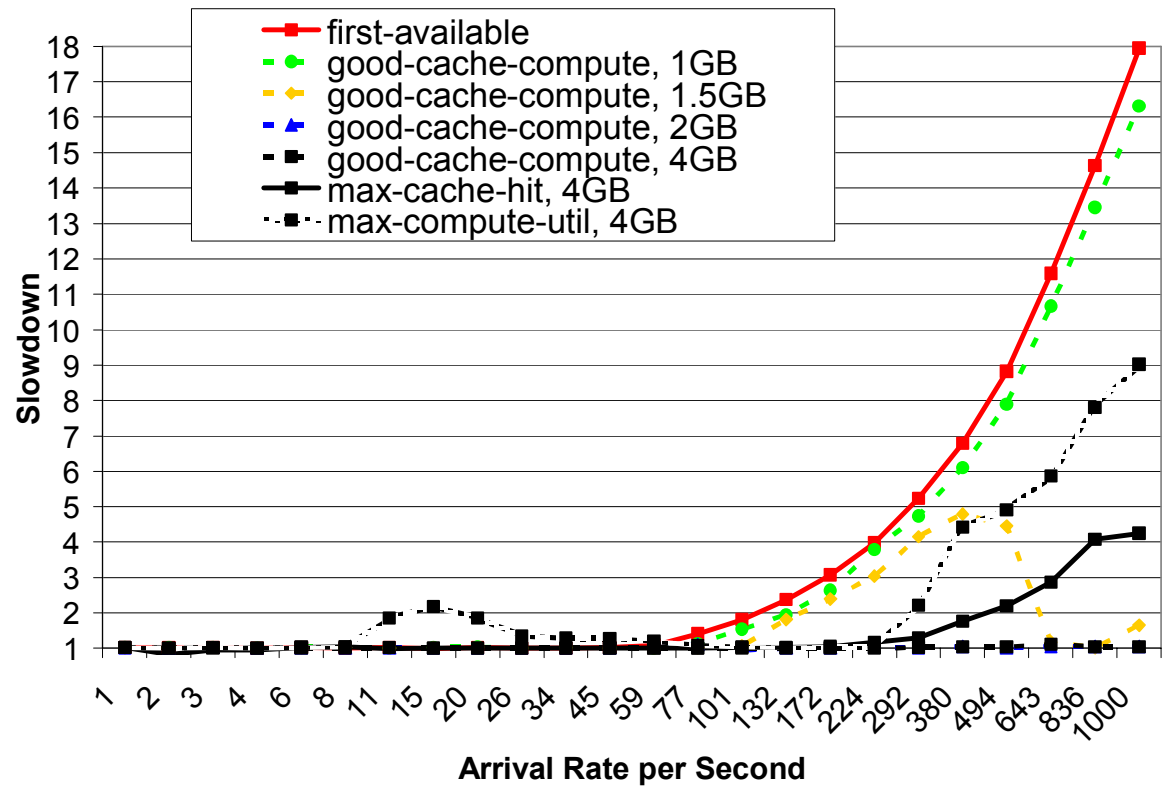

***Figure 14: Slowdown for the LAN experiment as we varied arrival rate***

It is important to note that resource allocation takes on the order of 30~60 seconds due to LRM's overheads, which is why it took the slowdown 120 seconds to return back to the normal (1X), as the dynamic resource provisioning compensated for the drop in performance.

### *5.2.6 Response Time*

The response time is probably one of the most important metrics from an application's point of view, as it determines if interactivity is plausible for a given workload, and can influence the performance perception of the resource management and the particular set of resources used. *Average Response*

*Time ($AR_T$)* is the end-to-end time from task submission to task completion notification; $AR_T = WQ_T+E_T+D_T$, where $AR_T$ is the average response time, $WQ_T$ is the wait queue time, $E_T$ is the task execution time, and $D_T$ is the delivery time to deliver the result.

Figure 15 shows response time results across all 14 experiments in log scale. We see a significant different between the best data diffusion response time (3.1 seconds per task) to the worst data diffusion (1084 seconds) and the worst GPFS (1870 seconds).

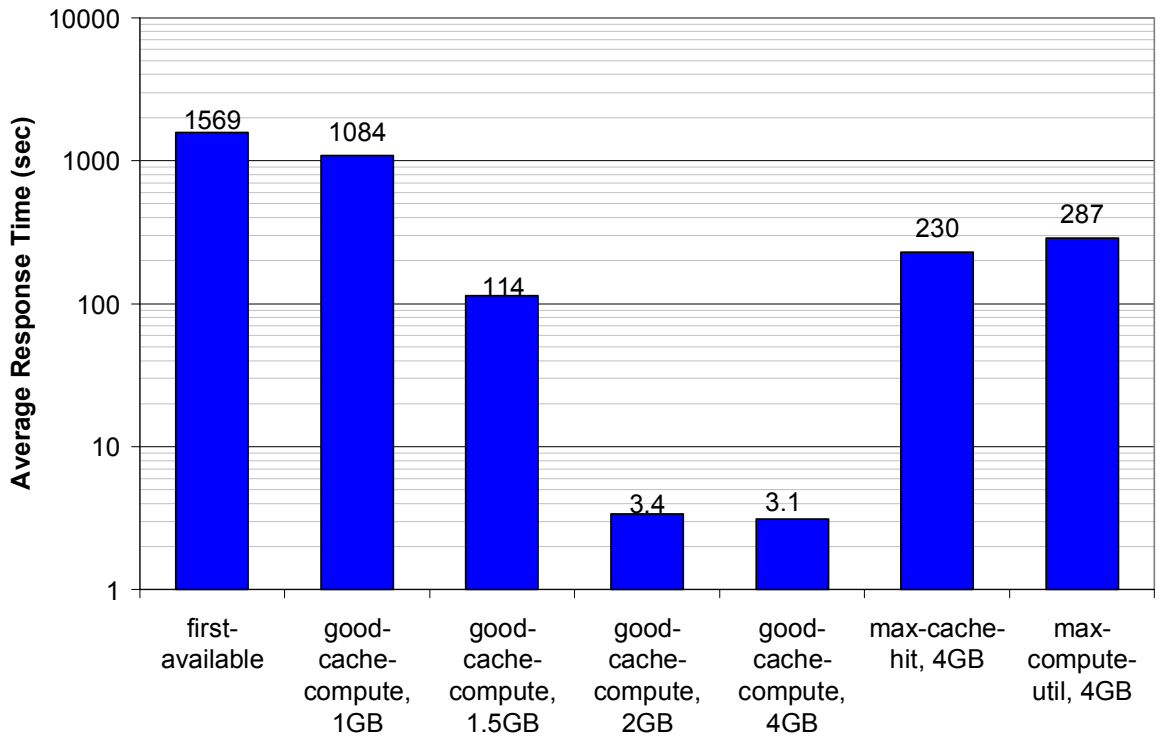

***Figure 15: Average response time for LAN and WAN***

That is over 500X difference between the data diffusion good-cache-compute policy and the first-available policy (GPFS only) response time. One of the main factors that influences the average response time is the time tasks spend in the Falkon wait queue. In the worst (first-available) case, the queue length grew to over 200K tasks as the allocated resources could not keep up with the arrival rate. In contrast, the best (good-cache-compute with 4GB caches) case only queued up 7K tasks at its peak. The ability of the data diffusion to keep the wait queue short allowed it to achieve an average response time of only 3.1 seconds.

# 6. CONCLUSIONS

Dynamic analysis of large datasets is becoming increasingly important in many domains. When building systems to perform such analyses, we face difficult tradeoffs. Do we dedicate computing and storage resources to analysis tasks, enabling rapid data access but wasting resources when analysis is not being performed? Or do we move data to compute resources, incurring potentially expensive data transfer costs?

We describe here a *data diffusion* approach to this problem that seeks to combine elements of both dedicated and on-demand approaches. The key idea is that we respond to demands for data analysis by allocating data and compute systems and migrating code and data to those systems. We then retain these dynamically allocated resources (and cached code and data) for some time, so that if workloads feature data locality, they will obtain the performance benefits of dedicated resources.

To explore this approach, we have extended the Falkon dynamic resource provisioning and task dispatch system to cache data at executors and incorporate data-aware scheduling policies at the dispatcher. In this way, we leverage the performance advantages of high-speed local disk and reduce access to persistent storage.

This paper has two contributions: 1) defining an abstract model for "data diffusion" and validating it against results from a real astronomy application; and 2) the exploration of the process of expanding a set of resources based on demand, and the impact it has on application performance. Our results show data diffusion offering dramatic improvements in performance achieved per resources used (34X) and that it reduces application response time by as much as 506X when compared with data-intensive benchmarks directly against a shared file system such as GPFS.

In future work, we plan to explore more sophisticated algorithms that address, for example, what happens when an executor is released; should we discard cached data, should it be moved to another executor, or should it be moved to persistent storage; do cache eviction policies affect cache hit ratio performance? Answers to these and other related questions will presumably depend on workload and system characteristics.

We plan to use the Swift parallel programming system to explore data diffusion performance with more applications and workloads. We have integrated Falkon into the Karajan workflow engine used by Swift [14, 28]. Thus, Karajan and Swift applications can use Falkon without modification. Swift has been applied to applications in the physical sciences, biological sciences, social sciences, humanities, computer science, and science education. We have already run large-scale applications (fMRI, Montage, MolDyn, DOCK, MARS) without data diffusion [4, 14, 28, 32], which we plan to pursue as use cases for data diffusion.

# 7. ACKNOWLEDGEMENTS

This work was supported in part by the NASA Ames Research Center GSRP Grant Number NNA06CB89H and by the Mathematical, Information, and Computational Sciences Division subprogram of the Office of Advanced Scientific Computing Research, Office of Science, U.S. Dept. of Energy, under Contract DE-AC02-06CH11357. We also thank TeraGrid and the Computation Institute at University of Chicago for hosting the experiments reported in this paper.